# Ultimate In-plane Magnetoresistance Ratio of Graphene by Controlling the Gapped Dirac Cone through Pseudospin

*Yusuf Wicaksono\*[1], Halimah Harfah[1], Gagus Ketut Sunnardianto[2], Muhammad Aziz Majidi[3], Koichi Kusakabe[1]*

1. Graduate School of Engineering Science, Osaka University, 1-3 Machikaneyama-cho, Toyonaka, Osaka 560-8531, Japan

2. Research Center for Physics, Indonesian Institute of Sciences, Kawasan Puspiptek Serpong, Tangerang Selatan, 15314, Banten, Indonesia

3. Department of Physics, Faculty of Mathematics and Natural Science, Universitas Indonesia, Kampus UI Depok, Depok, Jawa Barat 16424, Indonesia

ABSTRACT. A theoretical study is presented on the in-plane conductance of graphene that is partially sandwiched by Ni(111) slabs with a finite size and atom-scale width of ~12.08 Å. In the sandwiched part, the gapped Dirac cone of graphene can be controlled via pseudospin by changing the magnetic alignment of the Ni(111) slabs. When the magnetic moments of the upper and lower

Ni(111) slabs have antiparallel and parallel configurations, the bandgap at the Dirac cone is open and closed, respectively. The transmission probability calculation for the in-plane conductance of the system indicated that the antiparallel configuration would result in nearly zero conductance of $E - E_F = 0.2$ eV. In the parallel configuration, the transmission probability calculation indicated that the system would have a profile similar to that of pristine graphene. A comparison of the transmission probabilities of the antiparallel and parallel configurations indicated that a high magnetoresistance of 1450% could be achieved. An ultimate magnetoresistance can be expected if the Ni(111) slab widths are increased to the nanometer scale.

The extraordinary in-plane charge mobility of graphene[1] makes it an ideal material for applications in microelectronics[2,3] and sensing[4] owing to its novel electronic structure. The equipotential of C atoms in sublattices A and B of the graphene layer causes a Dirac cone energy band at zero energy. This Dirac cone energy band causes the electrons in graphene to behave in a peculiar way, where they all have the same velocity and absolutely no inertia[2]. Growing graphene on top of a transition metal substrate enables sensitive tuning of the Dirac cone energy band. When graphene is placed on a metal substrate, hybridization might occur, which breaks the chiral symmetry of graphene and creates a gapped Dirac cone[5]. Another unique property of graphene is its magnetic response when grown on top of a ferromagnetic substrate[6-8]. This induces a magnetic moment in the graphene because of charge transfer and has an orientation opposite to that of the ferromagnetic substrate[8]. Additionally, the weak spin–orbit coupling and long spin scattering length of graphene[9] make graphene a prospective material for spintronic devices. Many studies have attempted to fabricate graphene-based spintronic devices for logic devices, hard-drive magnetic read heads, and magnetic sensors[10-15]. Ni(111) surfaces are the most commonly used contacts to study graphene-based spintronic devices owing to their similar structure (smallest lattice mismatch among transition metals) and strong hybridization with graphene[7-8,16]. Graphene has been used as a bridge between two Ni electrodes[17-19] because of its long spin relaxation lifetime[9] or as a tunnel barrier in a Ni/graphene/Ni magnetic tunnel junction (MTJ)[20-30]. Although further results are expected, a new strategy for achieving a low resistance and overcoming the absence of a bandgap is necessary to realize the application of graphene in spintronic devices.

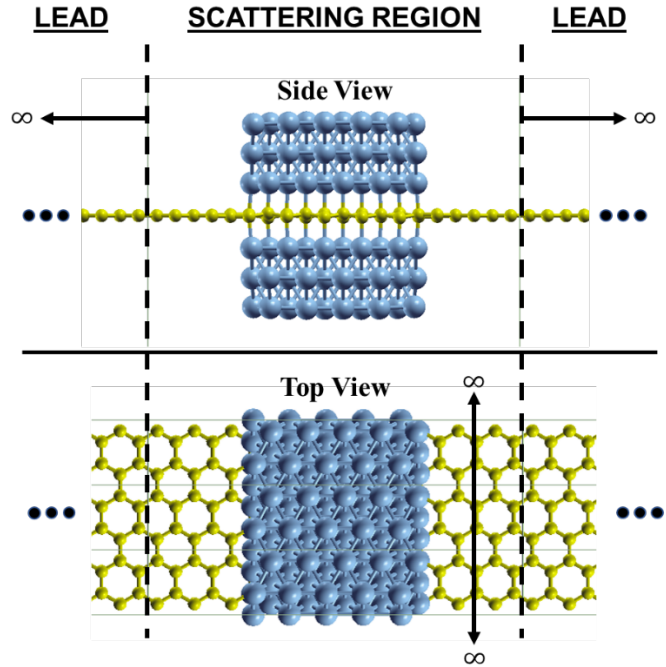

**Figure 1**. Proposed system of graphene where the middle part is sandwiched by Ni(111) slabs. The most stable stacking arrangement from our previous study was considered. The graphene is used as a buffer and electrode.

Our previous study showed that, for a Ni/graphene/Ni heterostructure, the gapped Dirac cone of graphene can be controlled by the magnetic alignment of the Ni slabs[31]. When the magnetic moments of the upper and lower Ni(111) slabs have an antiparallel configuration (APC), the bandgap at the Dirac cone is open. In a parallel configuration (PC), the bandgap is closed. This unique characteristic is because the most stable arrangement of the Ni/graphene/Ni heterostructure occurs when Ni atoms of the upper and lower Ni(111) slabs at the interfaces hybridize with the different sublattices of graphene. In other words, Ni atoms from the lower Ni(111) slab hybridize with C atoms in sublattice A (i.e., $C_A$), and Ni atoms from the upper Ni(111) slab hybridize with C atoms in sublattice B (i.e., $C_B$). Although the C atoms of graphene bond with Ni atoms, this special hybridization preserves the equipotential between sublattices A and B. Meanwhile, a magnetic moment is induced on the graphene layer by charge transfer from the Ni to C atoms. The induced magnetic moments of the $C_A$ and $C_B$ atoms depend on the magnetic alignment of the Ni(111) slabs. In APC and PC, the induced magnetic moments between the $C_A$ and $C_B$ atoms have antiferromagnetic and ferromagnetic orders, respectively. This means that the pseudospin between sublattices A and B can be controlled to preserve or break the chiral symmetry, so the gap of the Dirac cone is also controllable. A controllable gapped Dirac cone should help realize a high in-plane magnetoresistance (MR) for graphene. This paper presents a theoretical study on the in-plane conductance of graphene partially sandwiched with Ni(111) slabs. The aim was to investigate the effectiveness of a controllable Dirac cone on the conductance of graphene to realize a high MR ratio.

Ni(111) slabs with a finite size and atom-scale width of ~12.08 Å were considered, as shown in **Figure 1**. Graphene was used as the buffer layer and the electrodes in the calculation. The same stacking arrangement used in the previous study was considered, where Ni atoms at the interfaces of the lower and upper Ni(111) slabs hybridize with C atoms in sublattices A and B, respectively. Both APC and PC were considered for the Ni(111) slabs. Spin-polarized plane wave-based density functional theory (DFT) calculations were performed using the Quantum ESPRESSO package[32-33] to obtain the structural equilibrium and spin-charge density properties of the proposed system. The electron–ion interaction was described by a revised Perdew–Burke–Ernzerhof (PBE) functional for a densely packed solid surface (i.e., PBESol functional)[34] and ultrasoft pseudopotentials[35] within the generalized gradient approximation (GGA). The atomic positions were relaxed with a total force tolerance of 0.001 eV/Å. A 45 × 45 × 1 Monkhorst–Pack *k*-mesh was used for the calculations. First-principles quantum transport calculations, which couple DFT with the non-equilibrium Green's function, were performed with the Siesta and Transiesta packages[36-39] to calculate the transport properties at a zero-bias voltage. The PBESol functional and Troullier–Martins pseudopotential[40] were used within the GGA. The double-zeta plus polarization basis set[41-43] was employed, and the temperature was set to 300 K. A perpendicular *k*-point of 1 × 901 with respect to the transmission direction was considered to obtain good accuracy for the transmission probability.

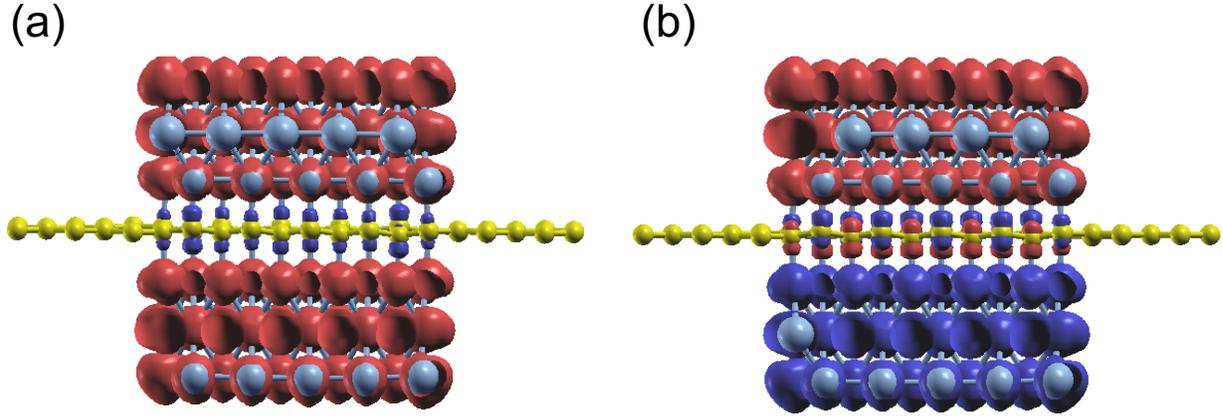

**Figure 2.** Spin-charge density mapping of the proposed system when the Ni(111) slabs are in (a) PC and (b) APC (red: spin-up charge density; blue: spin-down charge density).

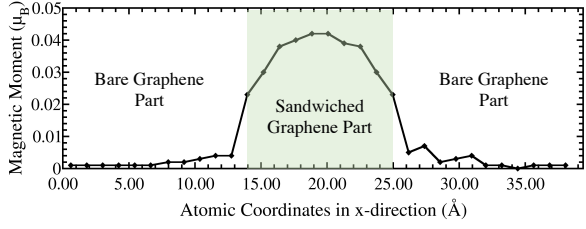
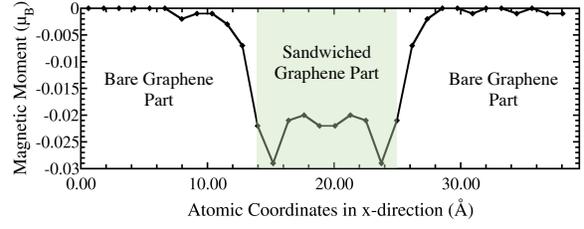
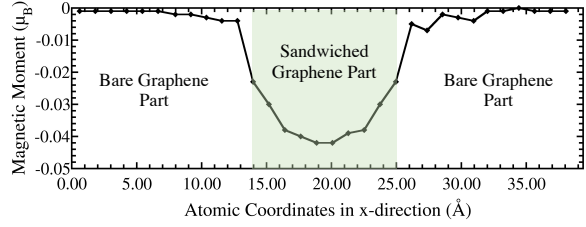
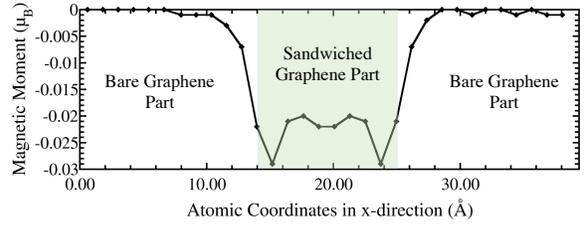

*Figure 3.* *Induced magnetic moment on graphene: (a) APC and (b) PC.*

The effect of the induced magnetic moment on graphene was investigated by considering the finite size of the Ni(111) slabs. Particular focus was placed on the boundaries between bare and sandwiched graphene. Mapping of the spin-charge density, as shown in Figure 2, indicated that the magnetic moment was only induced in graphene sandwiched by Ni(111) slabs up to the boundaries. However, the details of the induced magnetic moment profile, which is shown in Figure 3, indicate that the magnetic moment is damped to a small value on the bare graphene part near the boundaries for a few C atoms before reaching zero. This damping corresponds to wave function matching at the boundaries between the sandwiched and bare graphene.

The C atoms of bare graphene far from the boundaries do not have an induced magnetic moment, as shown in Figure 2 and Figure **3**. In this case, the Dirac cone of graphene is similar to that of pristine graphene because the chiral symmetry is preserved: $n_{A\sigma} = n_{B\sigma}$ with $\sigma = \uparrow$ or $\downarrow$. Meanwhile, the sandwiched graphene has an induced magnetic moment depending on the magnetic alignment of the Ni(111) slabs. For Ni(111) slabs in APC, Figure 2(a) and Figure 3(a) indicate that the induced magnetic moments on the $C_A$ and $C_B$ atoms of the sandwiched graphene have an antiferromagnetic order. Thus,

$$n_{A\downarrow} = n_{B\uparrow} > n_{A\uparrow} = n_{B\downarrow}, \tag{1}$$

which implies that the chiral symmetry is broken, and the gap of the Dirac cone is open.

When the Ni(111) slabs are in PC, the induced magnetic moments in the $C_A$ and $C_B$ atoms have the same magnetic orientation and almost the same value, as shown in Figure 2(b) and Figure **3**(b). Thus,

$$n_{A\downarrow} = n_{B\downarrow} > n_{A\uparrow} = n_{B\uparrow}. \tag{2}$$

This means that the equipotential between $C_A$ and $C_B$ preserves the chiral symmetry of graphene, and the Dirac cone survives. However, because of the induced magnetic moment on the sandwiched graphene, the Dirac cone on the spin majority channel is lower than that on the spin minority channel, which is characteristic of ferromagnetic materials.

The magnetic moment characteristics of the $C_A$ and $C_B$ atoms in graphene agree with the findings of a previous study[31]. Interestingly, the characteristics of the induced magnetic moments in the sandwiched graphene part were different for APC and PC. In APC, the amplitude of the induced magnetic moment decreases from the center of the sandwiched graphene along the direction of the boundary. The magnetic moment amplitude of C atoms at the center of the sandwiched graphene is in agreement with the magnetic moment of the C atoms of Ni/Gr/Ni in a periodic system (~0.04 $\mu_B$). This means that the decrease in the magnetic moment amplitude along the direction to the boundary corresponds to the characteristics that are found in the finite size of the Ni slabs.

Based on the structure of the system, this decrease might be due to the coordination of Ni(111) slabs, which compress their shape toward the center but maintain their bond with C atoms at the interfaces. An edged shape forms on Ni(111) slabs at the boundaries, which increases the lattice mismatch between the Ni layer at the interfaces and the graphene layer, and in turn, decreases the charge transfer from Ni atoms to C atoms. Thus, the reduced charge transfer from Ni atoms to C atoms at the interface decreases the induced magnetic moment on the C atoms. Because the lattice mismatch increases along the direction of the boundary, the induced magnetic moment also decreases along the same direction. In PC, the profile of the induced magnetic moment on the sandwiched graphene is unique. The induced magnetic moment initially decreased from the center of the sandwich to the boundaries. However, it then starts to increase to become even higher than that of the center part before finally decreasing again near the boundaries. The magnetic moment of the C atoms in the center of the sandwiched graphene is approximately equal to the induced magnetic moment of the C atoms of the Ni/Gr/Ni system in the periodic system, which is ~0.022 $\mu_B$. Furthermore, the decrease in the induced magnetic moment on the neighboring C atoms, which shifted from the center of the sandwiched to the boundary, corresponds to the increasing lattice mismatch, similar to the case of APC. Just before at the boundaries, the amplitude of the magnetic moment on the C atoms increases even higher than that of in the center part. This might be due to the reduction in the antiferromagnetic configuration between sublattices A and B near the boundary. In the periodic system of Ni/Gr/Ni, the amplitude of the induced magnetic moment of C atoms in APC is higher than that of PC[31], which is due to the fact that the carbon atoms of sublattices A and B in graphene tend to have an antiferromagnetic (AFM) configuration because of the half-filled $p_z$-orbital and Pauli's exclusion principle. This rule is often found in organic molecules in $sp^2$-hybridization or magnetic alternant hydrocarbon systems[44-45]. Near the boundaries, the contribution of the antiferromagnetic alignment between the hybridized C and Ni atoms is more dominant than that of the antiferromagnetic alignment between C atom sublattices A and B. Thus, it leads to an increase in the induced magnetic moment. Finally, at the boundaries, the induced magnetic moment of the C atoms in the sandwiched part decreases again, which indicates the effect of increasing the lattice mismatch.

These changes in the induced magnetic moment on the sandwiched graphene affect the Dirac cone characteristics. When the Ni(111) slabs are in APC, decreasing the induced magnetic moment reduces the size of the gap of the Dirac cone. This means that the Dirac cone gap is smaller at the boundaries of the sandwiched graphene than at the center. For Ni(111) slabs in PC, the changes in the induced magnetic moment affect the size of the Stoner gap between the spin majority channel and spin minority channel.

Although the sandwiched graphene part is expected to result in a controllable Dirac cone, it is also expected to shift the Dirac cone to a higher energy position. Thus, the sandwiched graphene can act as a potential barrier between two bare graphene layers. When electrons are transmitted through this potential barrier of the sandwiched graphene, opening and closing the Dirac cone gap may not significantly affect the in-plane conductance of graphene. Therefore, the zero-bias limit was considered in the calculations of the transmission probability of electrons to investigate the effectiveness of opening and closing the gap of the Dirac cone of graphene on the in-plane conductance. Figure 4 shows the transmission probability of electrons through sandwiched graphene.

When the Ni(111) slabs are in PC, the transmission probability of the proposed system has a profile similar to that of pristine graphene, as shown in Figure 4(a). The typical transmission probability of pristine graphene comes from the Dirac cone of graphene, where the transmission probability results in zero conductance at the Fermi energy and increases linearly with energy. This transmission probability profile was also found at $E - E_F = 0.76$ and $0.82$ eV for the spin-up and spin-down electron transmission probabilities, respectively, which are denoted as A and B in Figure 4(a). The additional zero transmission probability, which comes from the Dirac cone at a higher energy, corresponds to the shifted and spin-polarized Dirac cones, respectively, of the sandwiched graphene. This agrees with the Ni/graphene/Ni band structure reported in a previous study. Interestingly, the potential barrier created on the sandwiched graphene does not decrease the transmission probability of pristine graphene at energies below the additional Dirac cone, but slightly increases it. This means that electrons tunnel through the potential barrier from the sandwiched graphene, similar to Klein tunneling. The slight increase in the total transmission probability comes from spin-up electrons that transmit not only through graphene but also through Ni(111) slabs in the sandwiched graphene part. Thus, the spin-up electrons have a higher transmission probability than spin-down electrons. This transmission path also explains the non-zero transmission probability for the Dirac cone that was observed for spin-up electrons.

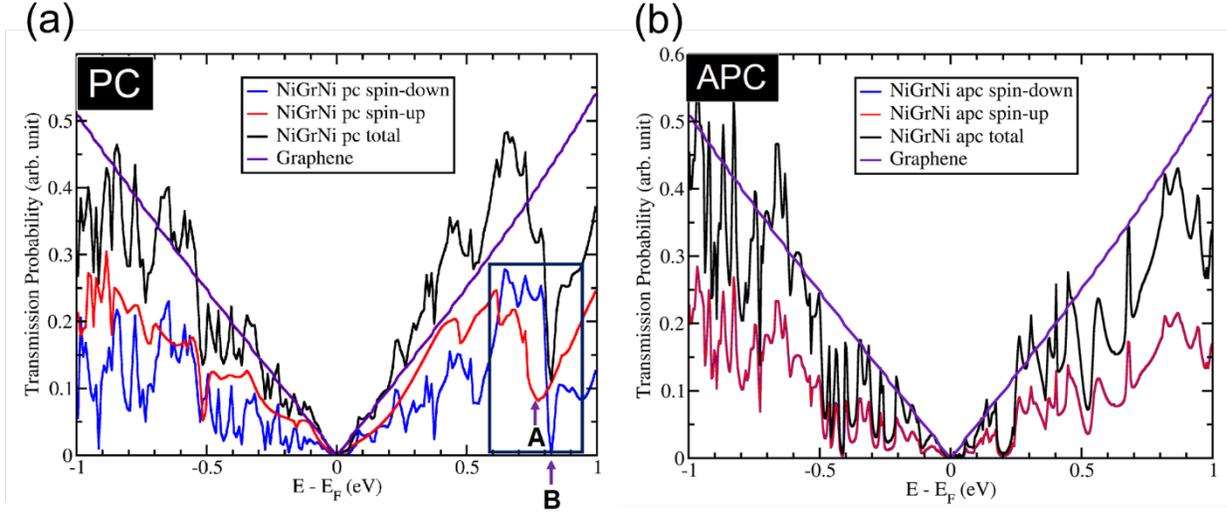

**Figure 4.** *Transmission probability of the proposed system depending on the magnetic alignment of the Ni(111) slabs: (a) PC and (b) APC.*

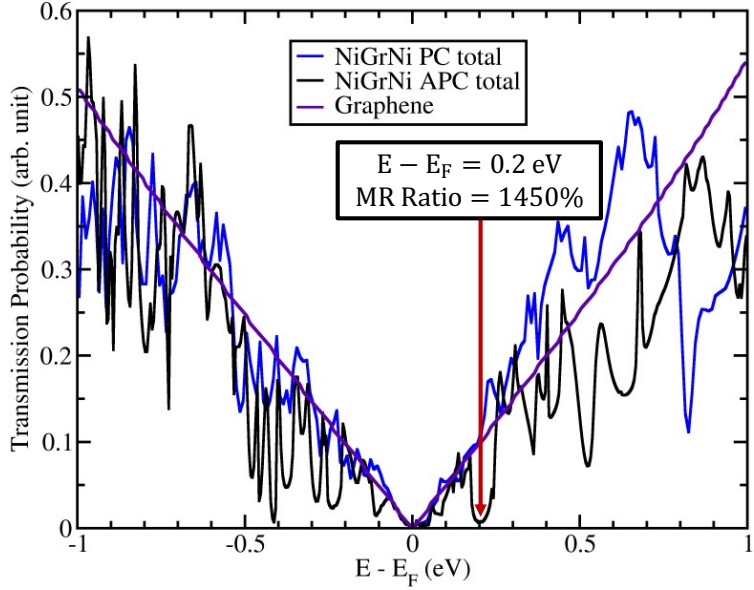

**Figure 5.** *Transmission probabilities of the proposed system with Ni(111) slabs in APC and PC. A high in-plane MR ratio of up to 1450% was found at $E - E_F = 0.2$ eV.*

A profile similar to that of pristine graphene was also observed when the Ni (111) slabs were in APC. However, a unique gap-like zero transmission probability for both spin-up and spin-down electrons was found at $E - E_F = 0.18 - 0.22$ eV, with the lowest transmission probability at $E - E_F = 0.2$ eV, as shown in Figure 4(b). In APC, the spin-up and spin-down electron transmissions overlap because of the condition in Equation (1). The gap-like zero transmission probability corresponds to the Dirac cone gap found in the Ni/graphene/Ni band structure in APC. However, this transmission probability shows that the zero transmission due to the gapped Dirac cone does not clearly correspond to the zero transmission probability. This is because the induced magnetic moment differs between the C atoms at the boundaries and the center of the sandwiched graphene. This implies that, when a sufficiently wide Ni(111) slab is considered (at the nanometer scale), perfect zero transmission can be expected depending on the size of the Dirac cone gap in the band structure. The nearly zero transmission implies that opening the gap of the Dirac cone causes the sandwiched graphene to act more like an insulator. The nearly linear increase in the transmission probability with increasing energy before and after the gap-like zero transmission indicates that the electrons tunnel through the potential barrier for the sandwiched graphene. The slight decrease in the transmission probability beside the gap-like zero transmission can be attributed to the difficulty of electron transmission from the spin-blockage of different magnetic orientations on the upper and lower Ni(111) slabs. The transmission probabilities of the proposed system with Ni(111) slabs were compared in PC and APC, and a high in-plane MR ratio of up to 1450% was observed at $E - E_F = 0.2$ eV, as shown in Figure 5. The use of a gate voltage to achieve a high in-plane MR ratio should shift the Fermi energy. Interestingly, this high in-plane MR ratio can only be achieved by sandwiching graphene with Ni(111) slabs with an atom-scale width (~12.08 Å). By increasing the slab width to the nanometer scale, the gap-like zero transmission probability for Ni(111) slabs in APC should have a transmission probability near absolute zero, and the high transmission should be maintained or even increased when the Ni(111) slabs are in PC. In this condition, the ultimate in-plane MR can be achieved with an MR ratio near an infinite value.

In summary, the induced magnetic moment of graphene sandwiched between Ni(111) slabs with a finite size and atom-scale width can be tuned. A slight decrease in the induced magnetic moment of C atoms near the boundary was observed, which corresponds to the compression of Ni(111) slabs at the edge and affects the charge transfer from Ni atoms at the interface to C atoms. The gap of the Dirac cone can be controlled according to the magnetic alignment of the Ni(111) slabs, which can result in a high in-plane MR ratio for graphene. A high in-plane MR ratio of up to 1450% was found considering only finite Ni(111) slabs with a width of ~12.08 Å. An ultimate in-plane MR ratio is expected when the Ni(111) slab width is increased to the nanometer scale.

AUTHOR INFORMATION

**Corresponding Author**

* (Y.W.) Email: wicaksono.y@opt.mp.es.osaka-u.ac.jp

**Author Contributions**

Conceptualization was performed by Y. W. K. K. supervised the research. Y.W. determined the equilibrium structure of the proposed system and performed spin-charge density mapping and transmission probability on the proposed system. Y.W. wrote the original draft of the manuscript. All authors have reviewed the manuscript. The manuscript was written through the contributions of all the authors. All authors approved the final version of the manuscript.

**Notes**

There are no conflicts of interest to declare.

ACKNOWLEDGMENT

Calculations were performed at the computer centers of Kyushu University. This work was partly supported by JSPS KAKENHI (Grant No. JP26400357, JP16H00914 in the Science of Atomic Layers, JP18K03456, and 20J22909 in a Grant-in-Aid for Young Scientists. Y. W. gratefully acknowledges the fellowship support from the Japan Society for the Promotion of Science (JSPS). H. H. gratefully acknowledges the scholarship support from the Japan International Cooperation Agency within the "Innovative Asia" Program (ID Number D1805252).